# Toward the QCD $\beta$ Function with Dynamical Wilson Fermions[*]

K.M. Bitar, R.G. Edwards, U.M. Heller, A.D. Kennedy and P. Vranas[a]

[a]Supercomputer Computations Research Institute Florida State University Tallahassee Florida 32306-4052, U.S.A.

We present data for the scaling behavior of lattice QCD with two flavors of light Wilson fermions. This is done by matching $\pi$ n and $\rho$ masses at the two lattice sizes of $16^3 \times 32$ and $8^3 \times 16$. We find that at $\beta = 6/g^2 \equiv 5.3$ the matching does not occur over a range extending down to $\beta = 3.5$. For $\beta = 5.5$ matching may be achieved at $\beta = 4.8 - 4.9$, leading to a $\Delta\beta = 0.6 - 0.7$ which is higher than the perturbative 2-loop value of 0.45. In both cases we conclude that the simulations are very far from the perturbative scaling region.

## 1. Introduction

If we measure some physical observable $\Omega$, such as a hadron mass, in a Monte Carlo computation of lattice QCD the result will not depend on the details of the lattice discretization as long as the lattice spacing $a$ is small enough. This is just the *assumption* that there is a universal continuum limit of lattice QCD. The value of the observable obtained by numerical calculation $\Omega = a^\delta f(g, \kappa)$ depends only on the parameters appearing in the lattice action, the gauge coupling $\beta = 6/g^2$ and the hopping parameter $\kappa$. The dimensionless quantity $f$ has been related to the dimensionful observable $\Omega$ by multiplying by the appropriate power of the lattice spacing $a$. As we approach the continuum limit $a \to 0$ the fact that $\Omega$ is independent of $a$ is expressed by the renormalization group equation

$$a\frac{d\Omega}{da} = a^\delta \left( \delta f - \beta_1 \frac{\partial f}{\partial g} - \beta_2 \frac{\partial f}{\partial \kappa} \right) = 0 \qquad (1)$$

where

$$\beta_1 = -a\frac{dg}{da}, \qquad \beta_2 = -a\frac{d\kappa}{da}. \qquad (2)$$

As we have two independent relevant parameters we require two independent physical observables to use as renormalization conditions to fix these parameters.

In a perturbative expansion of $\beta_1$ about $g = 0$ only the first two terms are independent of the details of the lattice regularization scheme.

The aim of this work is to determine how far the values of $g$ being used are from those at which $\beta_1$ is perturbative for the realistic case of QCD with two flavors of light Wilson fermions.

If we select some set of physical observables $\Omega_i$ we have

$$\Omega_i = a^{\delta_i} f_i(g, \kappa), \qquad (3)$$

where $\delta_i$ is the dimension of $\Omega_i$ and $f_i$ is the corresponding dimensionless function measured on the lattice. In the continuum limit all these observables must become independent of the lattice, so for two different lattice spacings $a_1$ and $a_2$

$$f_i\Big(g(a_1), \kappa(a_1)\Big) a_1^{\delta_i} = f_i\Big(g(a_2), \kappa(a_2)\Big) a_2^{\delta_i}. \qquad (4)$$

In practice we work with two lattices, $\Lambda_1$ with $L_1$ sites in each direction and with lattice spacing $a_1$, and $\Lambda_2$ with $L_2$ sites in each direction and lattice spacing $a_2$. By fiat we assert that $\Lambda_1$ and $\Lambda_2$ represent the same physical volume, thus

$$L_1 a_1 = L_2 a_2. \qquad (5)$$

Using this relation we recast Equation (4) in terms of dimensionless quantities only

$$f_i\Big(g(a_1), \kappa(a_1)\Big) \left(\frac{L_2}{L_1}\right)^{\delta_i} = f_i\Big(g(a_2), \kappa(a_2)\Big). \qquad (6)$$

We define the quantity

$$\Delta\beta \equiv \frac{6}{g^2(a_2)} - \frac{6}{g^2(a_1)} \qquad (7)$$

---

[*]Talk presented by K.M. Bitar at Lat94 Conference, September 27 – October 1, 1994, Bielefeld, Germany.

to be change in the coupling needed to compensate for this change in the cutoff.

## 2. The $\beta$-function

The $\beta$-function at fixed $\kappa$[1] is defined by $\beta_1(g) = -a\,dg/da$, and the two universal terms from perturbation theory are $\beta_1(g) = -b_0 g^3 - b_1 g^5 + \cdots$. The solution of these equations, expressed in terms of $\beta = 6/[g(a_2)]^2$, is

$$\ln\left(\frac{a_2}{a_1}\right) = \frac{\sqrt{6}}{2}\int_{\beta-\Delta\beta}^{\beta}\frac{dx}{x^{3/2}\beta_1\left(\sqrt{6/x}\right)}; \qquad (8)$$

after evaluating the integral we find that

$$\Delta\beta_{\mathrm{Pert}} \simeq -12\left(b_0 + \frac{6b_1}{\beta}\right)\ln\left(\frac{a_2}{a_1}\right) \qquad (9)$$

which is the perturbative prediction for the change in $\beta$ needed to effect a change in scale by a factor $a_2/a_1$.

## 3. Previous Results

The $\beta$ function for two flavors of Wilson fermions was computed by some of us earlier [1]. The physical observables used for that purpose were a $\pi$-like state and several Creutz ratios of Wilson loops.

The main conclusion then was that perturbative scaling is expected to occur only at $\beta > 6.0$. In those simulations the $\pi$-like object had a heavy mass and it was thought at the time that this may have hidden the effects of the dynamical fermions in the system. Simulations with lighter $\pi$ masses are thus necessary to clarify the issue.

## 4. Present Simulations

In this work we concentrated on data available on lattices of size $16^3 \times 32$ with two flavors of Wilson fermions with parameters and results shown in Table 1.

The values of $\beta$ were chosen for hadron spectrum and other calculations largely based on the assumption that the presence of two fermion flavors renormalizes $\beta_{\mathrm{quenched}}$ downwards by about 0.5 and thus we may not be far from the perturbative scaling regime at these parameter values.

---

[1] We expect the $\beta$-function to become independent of $\kappa$ in the continuum and chiral limits.

Table 1
Parameters and Masses on $16^3 \times 32$.

| $\beta$ | $\kappa$ | $am_\pi$ | $am_\rho$ | $m_\pi/m_\rho$ |
|---|---|---|---|---|
| 5.3 | 0.1670 | 0.4540(20) | 0.6350(20) | 0.7150(54) |
| 5.3 | 0.1675 | 0.3120(40) | 0.5230(40) | 0.597(15) |
| 5.5 | 0.1604 | 0.2615(31) | 0.4278(70) | 0.611(20) |

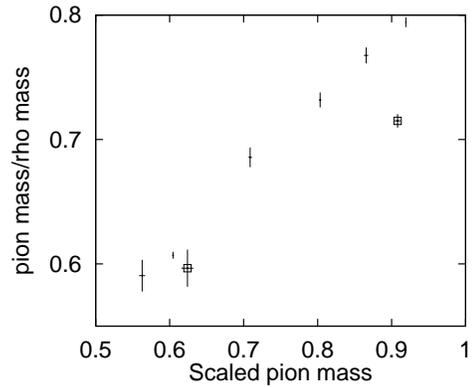

Figure 1. Crosses: $\beta = 3.5$ ($8^3 \times 16$); □: $\beta = 5.3$ ($16^3 \times 32$).

For our test of scaling we used two hadron masses ($m_\pi$ and $m_\rho$) as observables for matching with similar data generated on lattices of size $8^3 \times 16$ at values of $\beta$ ranging from 5.1 down to 3.5 and many $\kappa$ values.

## 5. Results

Our first result is the unmatchability of the data at $\beta = 5.3$. We find here that $\Delta\beta$ must be larger than 1.8. Figure 1 shows the two points obtained at $\beta = 5.3$ on the larger lattice (with $m_\pi$ appropriately scaled by a factor of 2) and the corresponding data for $m_\pi$ and $m_\pi/m_\rho$ on the smaller volume for $\beta \equiv 3.5$.

Our second result is that preliminary analysis of our data shows that at $\beta = 5.5$, matching is possible (Figure 2) with points at $\beta$ between 4.8 and 4.9 on the smaller lattice. We find for this case therefore that $\Delta\beta = 0.6$ to $0.7$.

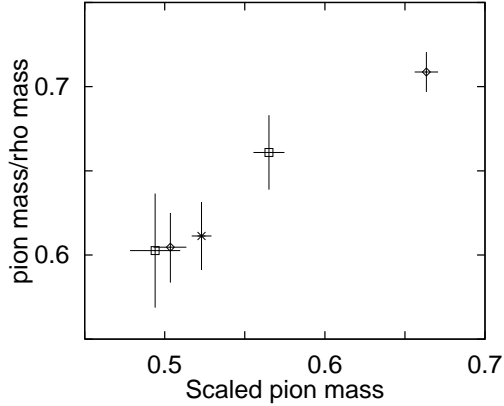

Figure 2. $*$: $\beta = 5.5$ ($16^3 \times 32$); $\diamond$: $\beta = 4.8$, o: $\beta = 4.9$ ($8^3 \times 16$).

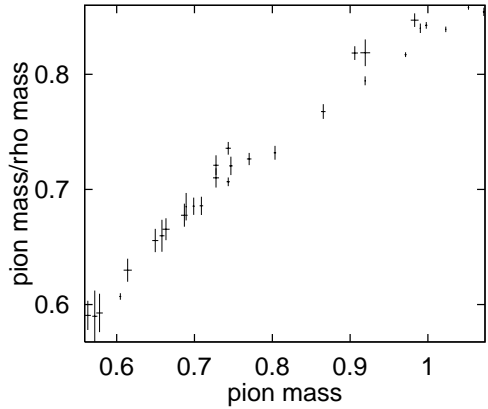

Figure 3. Cummulative data for $3.5 < \beta < 4.7$ ($8^3 \times 16$).

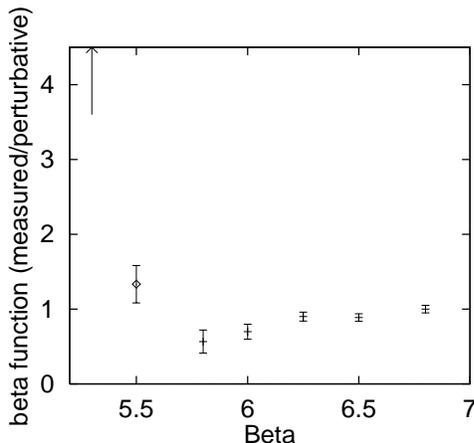

Figure 4. QCD $\beta$-function for $m_\pi/m_\rho \approx 0.6$ ($\diamond$) and from Ref. 1 ($-$).

A third result is an approximate parameter reduction in the theory at strong coupling: $\beta = 3.5$ to $\beta = 4.7$. Figure 3 includes all our data in this range of $\beta$ on the smaller lattice and shows that for $3.5 < \beta < 4.9$, the data is essentially a function of a single parameter. A change in $\beta$ may be compensated for by a change in $\kappa$.

## 6. Conclusions

Combining the present results with those obtained earlier on smaller lattices one is led to the conclusion that current dynamical simulations are being done at strong coupling. This is clear from the $\beta$-function we have measured, which is shown in Figure 4.

One therefore may need to reassess estimates for parameters and hence the cost of future computations with dynamical Wilson fermions accordingly.

## Acknowledgements

This research was supported by by the U.S. Department of Energy through Contract Nos. DE-FG05-92ER40742 and DE-FC05-85ER250000.## REFERENCES

1. K. Bitar, A. D. Kennedy and P. Rossi, *Phys. Rev. Lett.*, **63:25**, (1989), 2713.